\documentclass[]{spie}  

\usepackage{amsmath,amsfonts,amssymb}
\usepackage{graphicx}
\usepackage{siunitx}
\usepackage[colorlinks=true, allcolors=blue]{hyperref}

\title{Multi-scale second harmonic generation microscopy of ferroelectric domains in x-cut thin-film lithium niobate}

\author[a]{Sagar P. Doshi}
\author[a]{Gavin N. West}
\author[a, b]{Dodd Gray}
\author[a]{Rajeev J. Ram}
\affil[a]{Massachusetts Institute of Technology, 77 Massachusetts Ave, Cambridge, MA 02139, USA}
\affil[b]{MIT Lincoln Laboratory, 244 Wood Street, Lexington, MA 02421, USA}

\authorinfo{Further author information: (Send correspondence to S.P.D.)\\S.P.D.: E-mail: sdoshi@mit.edu}

\begin{document} 
\maketitle

\begin{abstract}
Thin-film lithium niobate (TFLN) is a widely used platform for nonlinear frequency conversion, as its strong nonlinear susceptibility and enhanced modal confinement intensify nonlinear interactions. Among phase matching techniques in TFLN, quasi-phase matching (QPM) is the dominant approach. For frequency doubling from near-infrared to visible wavelengths, this necessitates fabrication of QPM gratings with minimal period variation ($<$20nm) and control of ferroelectric domain inversion at the micron-scale along centimeter-long waveguides. Second harmonic generation microscopy (SHM) is a powerful tool for optimizing domain engineering (E-field poling), and it enabled the fabrication of near-ideal ($\sim$50\% duty cycle, $\sim$5\% variation in period) QPM gratings in \qty[mode=text]{5.6}{\mm}-long TFLN waveguides. Here, we show that increasing the SHM raster scan step size from \qtyrange[]{200}{400}{\nm} results in a 4x imaging speedup without sacrificing the accuracy of QPM grating characterization. To that end, we model device performance using Monte Carlo simulation of the coupled rate equations. Summary metrics of the QPM grating derived from SHM images are used as simulation inputs, and predictions of second harmonic output powers agreed well with experimentally measured values. We also modeled device performance using quasi-analytic correction factors, but we found that the predictions of second harmonic output power diverged from experimental measurements as the QPM grating quality degraded. To further speedup SHM imaging, we employed a statistical subsampling scheme that allows for characterization of \qty[mode = text]{5.6}{\mm}-long waveguides (poling period $\Lambda=3.240$) in approximately 5 minutes (30 seconds per field $\times$ 10 fields per waveguide). Each field is 100 microns in length, so our results indicate that sampling only $\sim$300 periods of a QPM grating is sufficient to accurately predict its second harmonic output. For the device characterized, this corresponds to $\sim$20\% of the total grating length. Together, discretization and device-length subsampling speed up SHM imaging by an order of magnitude. These results can enable wafer-scale imaging of TFLN devices, which is critical for realizing the scaling potential of this high-performance, integrated nonlinear platform. More generally, this work highlights the role of SHM as an invaluable tool for multi-scale materials characterization.

\end{abstract}


\section{INTRODUCTION}
\label{sec:intro}  
Thin-film lithium niobate (TFLN) is a widely used platform for nonlinear frequency conversion, because it combines large nonlinear coefficients and a multi-octave transparency window ($\sim$0.4-\qty[mode=text]{5.0}{\um}) with the enhanced modal confinement and massively-scalable fabrication capability of integrated photonics. Among phase matching techniques in TFLN, quasi-phase matching (QPM) is the dominant approach\cite{Zhu21}. Second harmonic generation (SHG) from near-infrared (NIR) to visible wavelengths necessitates the fabrication of QPM gratings with minimal period variation ($<$\qty[mode=text]{20}{\nm}) and control of ferroelectric domain inversion at the micron-scale along centimeter-long waveguides\cite{FejerQPM, Shur2015}. Second harmonic generation microscopy (SHM) is a powerful tool for evaluating QPM grating quality and optimizing domain engineering (E-field poling) of TFLN devices\cite{Zhao2020SubMicron, Zhao2020ShallowEtch, Jankowski2020UltrabroadbandNLO, Park22BlueSHG, Doshi2025}. A summary of the tradeoffs between SHM and other TFLN poling characterization modalities\cite{Reitzig2021} is shown in Tab. ~\ref{tab:CharacterizationModalities}. SHM has a number of key strengths that address the problem of logarithmic length scale characterization: it is non-destructive, can measure poling depth\cite{Rsing2019}, and acquires images 1-2 orders of magnitude faster than other imaging modalities\footnote{especially when factoring in sample preparation and measurement setup time (cleaning, chucking, focusing, etc.)} with only a moderate dropoff in imaging resolution. As such, we used it to optimize periodic poling of devices designed for NIR to visible SHG (poling period $\Lambda > \qty[mode = text]{1}{\um}$). 
\begin{table}[t]
    \caption{Summarizes the tradeoffs between SHM and other TFLN poling characterization modalities. Piezo force microscopy is abbreviated PFM, and HF etch+SEM indicates a hydrofluoric acid etch followed by scanning electron microscopy. The resolution limit for each of the modalities is instrument-dependent, so typical values\cite{Vladar2023, Kalinin2006, Benninger2013} are reported as ranges. The depth sensitivity of PFM is still under preliminary investigation \cite{Roeper2024}.} 
    \label{tab:CharacterizationModalities}
    \begin{center}       
    \begin{tabular}{|c|c|c|c|}
    \hline
    \rule[-1ex]{0pt}{3.5ex} & \textbf{HF Etch+SEM} & \textbf{PFM} & \textbf{SHM} \\
    \hline
    \rule[-1ex]{0pt}{3.5ex} Non-destructive? & X & \checkmark & \checkmark \\
    \hline
    \rule[-1ex]{0pt}{3.5ex} Resolution (\qty[mode = text]{}{\nm}) & $\sim$1-10 & $\sim$10-100 & $\sim$250-1000  \\
    \hline
    \rule[-1ex]{0pt}{3.5ex} Speed (\qtyproduct{100 x 40}{\um}) & $\sim$minutes-hours & $\sim$minutes-hours & $\sim$seconds  \\
    \hline
    \rule[-1ex]{0pt}{3.5ex}  Depth-sensitive? & \checkmark & $\sim$ & \checkmark  \\
    \hline 
    \end{tabular}
    \end{center}
\end{table}

A broader goal of recent works using the TFLN platform is to realize massively-scalable production of high-performance integrated nonlinear devices. To that end, the record second harmonic absolute conversion efficiency reported in the literature\cite{Chen2023} (80\%) required characterization of the TFLN thickness along the entire length of the \qty[]{2}{\cm}-long nonlinear waveguide. Moreover, there is continued work on scalable TFLN device fabrication including demonstration of $<$\qty[per-mode = symbol]{1}{\dB\per\cm} losses across an entire wafer\cite{Luke20} and wafer-scale periodic poling \cite{Chen2024}. Applying the fastest SHM imaging speed of TFLN reported in the literature\cite{Reitzig2021} ($\sim$1cm\textsuperscript{2}/ hour) to a 100mm wafer, it would take more than 2 days to image the $\sim$56 die of \qty[]{1}{cm^2} area in the wafer. This furthers the logarithmic length scale challenge of characterizing TFLN devices, so we sought methods to speed up SHM imaging without compromising its ability to evaluate device performance. Specifically, we demonstrate subsampling methods which speedup SHM imaging by an order of magnitude. Moreover, we develop numerical models which accurately predict device performance (second harmonic output power and conversion efficiency) with the subsampled SHM data.

\section{Methods}
\label{sec:meth}  

\begin{figure} [ht]
   \begin{center}
   \includegraphics[height=3.75cm]{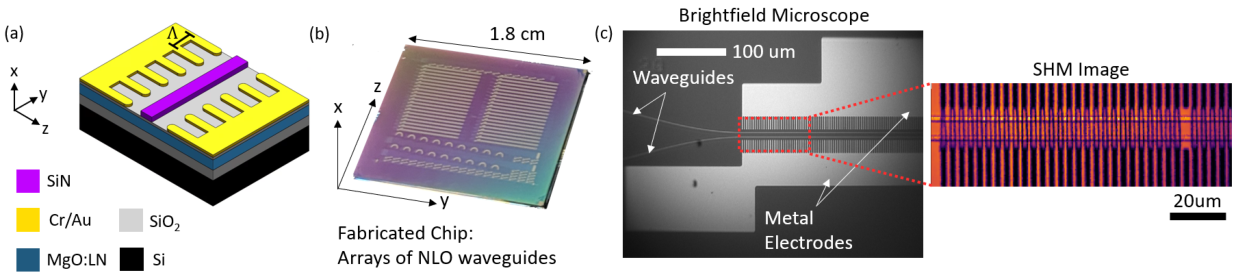}
   \end{center}
   \caption[example] 
    { \label{fig:DeviceStructure} 
    (a) sketches the device structure. For clarity, layer thicknesses and electrode dimensions are not drawn to scale. (b) Illustrates a photo of the fabricated chip with an array of nonlinear optical test structures, and (c) displays a brightfield image and corresponding SHM image of a device on the chip with a poling period of $\Lambda = $ \qty[]{3.240}{\um}.
    }
   \end{figure}

To access the largest element of the second order nonlinear tensor ($d_{33} = \frac{1}{2}\chi^{(2)}_{zzz}$) and minimize photorefractive damage \cite{Furukawa2000MgoPhotorefractive}, we used x-cut, MgO-doped TFLN. Device fabrication began by dicing a commercial 5\%MgO-doped 200nm TFLN wafer  (NanoLN, Jinan Jingzheng Electronics Co., Ltd.) into $\sim$\qty{1.5}{\cm^{2}} chips. The wafer had a \qty[mode = text]{2}{\um} buried oxide (SiO\textsubscript{2}) thickness and a Si handle. The remaining fabrication steps were done at the chip scale. Before patterning poling electrodes, a \qty{50}{\nm} layer of SiO\textsubscript{2} was deposited. This oxide barrier reduces the leakage current during application of high voltage poling pulses\cite{Nagy2019}, and we found that a low temperature PECVD process (\qty[mode = text]{100}{\degreeCelsius} vs. \qty[mode = text]{300}{\degreeCelsius}) reduced poling variability. Poling electrodes (90/\qty[mode = text]{10}{nm} Cr/Au) were patterned using electron beam lithography. The electrodes consisted of a dual finger-comb design with rounded tips for E-field enhancement\cite{Zhao2020ShallowEtch, Nagy2019}. The semicircular tips had a diameter equal to the electrode width, which was set to 40\% of the desired poling period $(\Lambda)$ to pre-compensate for domain broadening. The gap between electrode tips was \qty[mode = text]{15}{\um}, and the length of each finger from its tip to the bus electrode bar was \qty[mode = text]{20}{\um}. Finally, SiN strips were deposited to form hybrid SiN-TFLN  waveguides, and a sketch of the final structure is shown in Fig. \ref{fig:DeviceStructure}\textcolor{blue}{(a)}.

Prior to poling, fabricated chips were annealed at \qty[mode = text]{500}{\degreeCelsius} for 48 hours with O\textsubscript{2} flow in an anneal tube. This annealing step reduces the variation in spatial defect density and hence the duty cycle variation of inverted domains\cite{Doshi2025}. Iterative optimization of ferroelectric domain patterning was conducted by poling a series of test devices with a single parameter sweep and imaging the poled sequence of devices using SHM. Our chips had a Si handle, so we operated our SHM\cite{Doshi2025} in reflection mode\cite{Rsing2019}. Silicone oil dropped onto the chip's surface to prevent arcing during high voltage probing was cleaned off with acetone and isopropyl alcohol rinses followed by a 30 second sonication in heptane. Using feedback from the SHM images, the poling parameters were adjusted, and this cycle of poling, rapid solvent cleaning, and imaging was repeated until near-ideal QPM gratings were observed. The final poling parameters consisted of heating the chip to \qty{200}{\degreeCelsius} and applying a single high voltage pulse. High temperature poling was found to improve QPM grating quality via enhancement of defect mobility\cite{Doshi2025}, and the pulse parameters were: peak voltage $V_{pk} = $ \qty{475}{\volt}, hold time at peak voltage $t_{hold} = $ \qty{80}{\ms}, and ramp/fall rates of $500/100$ \unit[per-mode = symbol]{\volt\per\ms}. A SHM image of a device poled with these parameters (poling period $\Lambda = $ \qty{3.240}{\um}) is shown in Fig. \ref{fig:DeviceStructure}\textcolor{blue}{(c)}.

\begin{figure} [hb]
   \begin{center}
   \includegraphics[height=4.5cm]{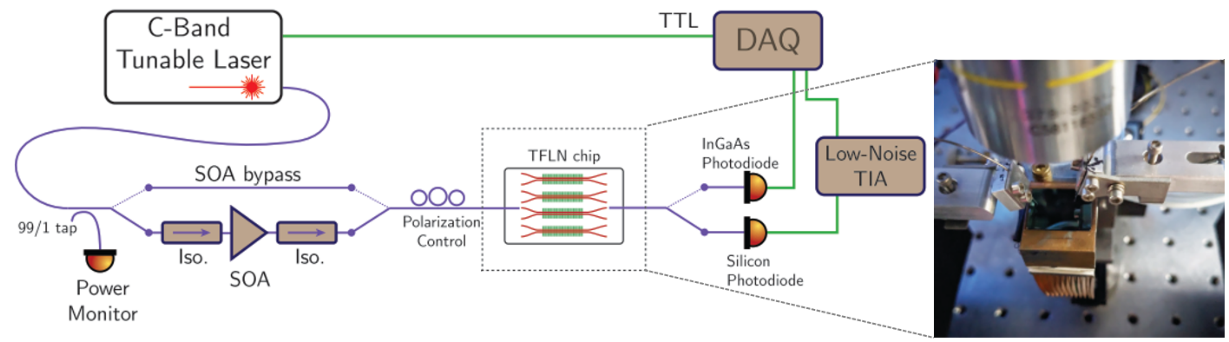}
   \end{center}
   \caption[example] 
    { \label{fig:OpticalCharacterizationSetup} 
    The experimental setup used to characterize waveguide devices. Isolators and the semiconductor optical amplifier are abbreviated iso. and SOA. An ultra-wideband tap and wavelength-calibrated power meter are used to monitor power launched into the amplifier (or the device, depending on configuration). For characterizing components at second harmonic wavelengths, a CW Ti:Sapphire laser providing NIR light (not shown) is coupled into fiber and routed through a wavelength-appropriate set of polarization control paddles and fiber probes. Transmitted powers are measured with photodetectors, whose output voltages are captured with a 16-bit data acquisition module (DAQ).
    }
   \end{figure}

After poling, linear and nonlinear optical characterization of TFLN devices was performed using the fiber probing setup shown in Fig. \ref{fig:OpticalCharacterizationSetup}. A tunable C-band laser provides pump light between \qtyrange{1480}{1583}{\nm}. A calibrated, ultra-broadband 99:1 fiber splitter and wavelength-calibrated power meter were used to normalize transmission measurements. For linear measurements, we bypass the SOA to avoid introducing spectral and current-dependent gain. This allows us to quickly measure the insertion loss of passive components. To characterize passive device behavior at second harmonic wavelengths, we use a CW Ti:Sapphire laser with a tuning range of roughly \qtyrange[range-phrase=--]{700}{1000}{\nm}. For nonlinear measurements, we routed light through the polarization-maintaining semiconductor optical amplifier (SOA) which provides $\sim$\qty[]{26}{\mW} of power with a 3dB bandwidth from \qty{1505}{\nm} to \qty{1583}{\nm}. Isolators are required on both ends of the SOA, as reflections from the vertical grating couplers can travel back into the fiber and induce oscillations. The contribution of amplified spontaneous emission (ASE) is calculated by comparing the transmitted power with and without an in-line, tunable narrowband filter. The filter is subsequently removed, because it introduces nontrivial ($\sim$2-3dB) loss. In short, this setup enables measurement of QPM spectra, coupling efficiencies, and conversion efficiencies.

\clearpage
\section{Results}
\label{sec:res}  
\begin{figure} [b]
   \begin{center}
   \includegraphics[height=8.5cm]{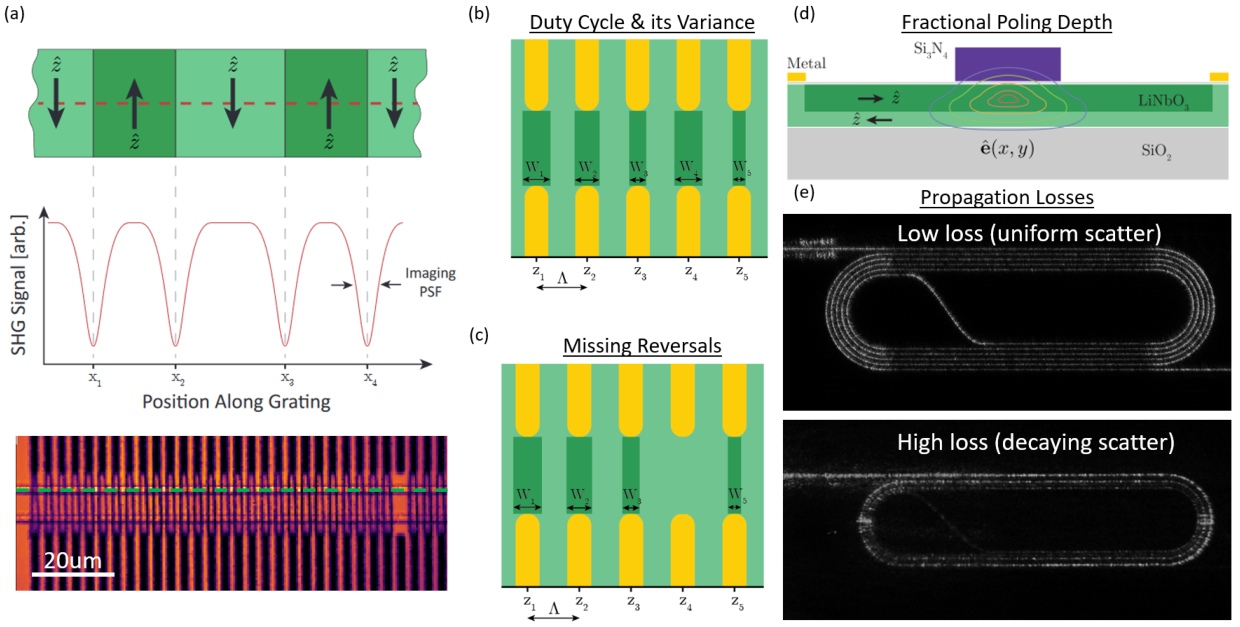}
   \end{center}
   \caption[example] 
    { \label{fig:nonIdealities} 
    (a) Illustrates the SHM image analysis process\cite{Doshi2025} (PSF = point spread function). The dashed green line overlaid on the SHM image highlights that the QPM grating was analyzed at the position of the SiN waveguide. (b) sketches variation in the duty cycle of each period where $DC_i = W_i/\Lambda$. In (c), there is a missing domain reversal in the 4th period. (d) overlays contour lines of a propagating mode on the cross section of a SiN-TFLN hybrid device with partial depth inversion. Finally, (e) shows two top-down brightfield microscope images of visible light propagating through a spiral delay line. The decay of the scatter is much more prominent in the high loss device.
    }
   \end{figure}

With the poled device shown in Fig. \ref{fig:DeviceStructure}\textcolor{blue}{(c)}, our goal was to develop and experimentally validate a numerical model for device performance. For SHG, a key empirical figure of merit is the length-normalized conversion efficiency $(\eta)$ in the undepleted pump approximation.
\begin{equation}
    \label{eq:convEff}
    \eta = \frac{P_{SH}}{L^2 P_{fund}^2} \times 100\% \ (\qty[per-mode = symbol]{}{\%.\watt^{-1}.\cm^{-2}})
\end{equation}
Here, $P_{SH}$ is the in-waveguide second harmonic output power, $P_{fund}$ is the in-waveguide fundamental (pump) power, and L is the nonlinear interaction length. Two numerical approaches were used to predict $\eta$ by analyzing SHM images of the QPM grating (Fig. \ref{fig:nonIdealities}\textcolor{blue}{(a)}). The first method was to extract QPM grating quality metrics and use them to calculate correction factors corresponding to each of the non-idealities illustrated in Fig. \ref{fig:nonIdealities}\textcolor{blue}{(b)-(e)}. The metrics needed to accurately predict $\eta$ are: the QPM grating mean duty cycle $(\mu_{DC})$, duty cycle standard deviation $(\sigma_{DC})$, probability of a missing domain reversal $(P_{mr})$, fractional poling depth $(f_{pol})$, and the device propagation losses (dB/cm) at the fundamental $(\alpha_{1})$ and second harmonic $(\alpha_{2})$.
Let $\eta_0$ be the conversion efficiency for a perfectly poled, lossless device. Then, the corrected conversion efficiency is given by Equation \ref{eq:eta_product}, which is the product of $\eta_0$ with correction factors for: deviation of the duty cycle mean from 50\% $(\eta_{DC})$, variation in the duty cycle $(\eta_{\sigma_{DC}})$, missing reversals $(\eta_{mr})$, propagation losses $(\eta_{prop})$, and a fractional poling depth $<$100\% of the TFLN thickness $(\eta_{f_{pol}})$.
\begin{equation}
    \label{eq:eta_product}
    \eta = \eta_0( \eta_{DC} \eta_{\sigma_{DC}} \eta_{mr}\eta_{f_{pol}}\eta_{prop})
\end{equation}
We set the fractional poling depth correction factor to unity $(\eta_{f_{pol}}=1)$, because SHM confirms domain inversion throughout the entire 200nm TFLN thickness\cite{Doshi2025, Rsing2019}. For $f_{pol} < 100\%$, the reduction in effective area $(A_{eff})$ of the nonlinear interaction can be used to calculate $\eta_{f_{pol}}$. The remaining correction factors are given by Equations \ref{eq:correction_factors}\textcolor{blue}{(a)-(d)}\cite{FejerQPM, Bortz1994LossCorrection}, where $l_c = \Lambda/2$ is the coherence length of the nonlinear interaction\cite{Boyd, Rsing2019} and $\sigma_{l} = \frac{\Lambda}{100}\sigma_{DC}$ is the domain length standard deviation.
\begin{subequations}
    \label{eq:correction_factors}
    \begin{align}
        \eta_{DC} &= \sin(\pi \mu_{DC})\\ 
        \eta_{\sigma_{DC}} &= 1-\frac{\pi^2 \sigma_{l}^{2}}{2l_{c}^{2}} \\
        \eta_{mr} &= (1-P_{mr})^2 \\
        \eta_{prop} &= \exp{[-2\alpha_{\omega}L]}\left( \frac{1-\exp{\left[-(\alpha_{2\omega}/2-\alpha_{\omega})L\right]}}{(\alpha_{2\omega}/2-\alpha_{\omega})L} \right)^2
\end{align}
\end{subequations}

The second approach to modeling device performance was to use SHM image data to define the second order nonlinear coefficient $\left( \chi^{(2)}(y)  \right)$ as a function of propagation distance\cite{Boyd}. The distance variable is $y$ (\qty[]{}{\cm}), as we are using x-cut TFLN with waveguides patterned along the crystal y-axis and NIR TE\textsubscript{00} pump light polarized along the crystal z-axis (Fig. \ref{fig:DeviceStructure}\textcolor{blue}{(a)}). We extend the in-waveguide, coupled wave-equations to account for small propagation losses $(\alpha_{i} \ll 2n_i\omega_i/c)$\cite{Chemla2012} and substitute $\chi^{(2)}(y)$ into the nonlinear coupling coefficients $\kappa_{i}$. For propagation losses similar to experimentally measured values ($\sim$\qty[per-mode = symbol]{1}{\dB\per\cm}), this set of differential equations uses the QPM grating structure embedded in $\kappa_{i}$ to predict a device's output second harmonic power. 
\begin{subequations}
    \label{eq:coupled}
    \begin{align}
        \frac{\partial A_1}{\partial y} &= j\omega_1 \kappa_1 A_{2}A_{1}^*e^{j\Delta ky} - \frac{\alpha_1}{2}A_{1} \\
        \frac{\partial A_{2}}{\partial y} &= j\omega_1\kappa_{2} A_{1}A_{1}e^{-j\Delta ky} - \frac{\alpha_{2}}{2}A_{1} \\
        \kappa_{i} &= \frac{\chi^{(2)}(y)}{cn_{i}A_{eff}}
    \end{align}
\end{subequations}
In Equations \ref{eq:coupled}\textcolor{blue}{(a,b)}, $n_i \ (i=1,2)$ are the effective indices of the fundamental and second harmonic waveguide modes. The phase mismatch is $\Delta k = 2k_{1}-k_{2} - 2\pi/\Lambda$, and $A_1$ and $A_{2}$ are the mode amplitudes of the fundamental and second harmonic waves respectively.  We numerically solve these equations using a Runge-Kutta pair method\cite{Tsitouras2011} implemented in the \texttt{DifferentialEquations.jl} package of the Julia programming language\cite{rackauckas2017differentialequations}. The equations are initialized with $A_{1}(y=0)=\sqrt{P_{fund}(y=0)}$ and $A_{2}(y=0)=0$. The solver discretization size is dynamically chosen by an internal adaptive time stepping algorithm based on specified local tolerance values, so we use a relative tolerance value of 1e-10 for stable convergence. 

\begin{figure} [th]
   \begin{center}
   \includegraphics[height=9cm]{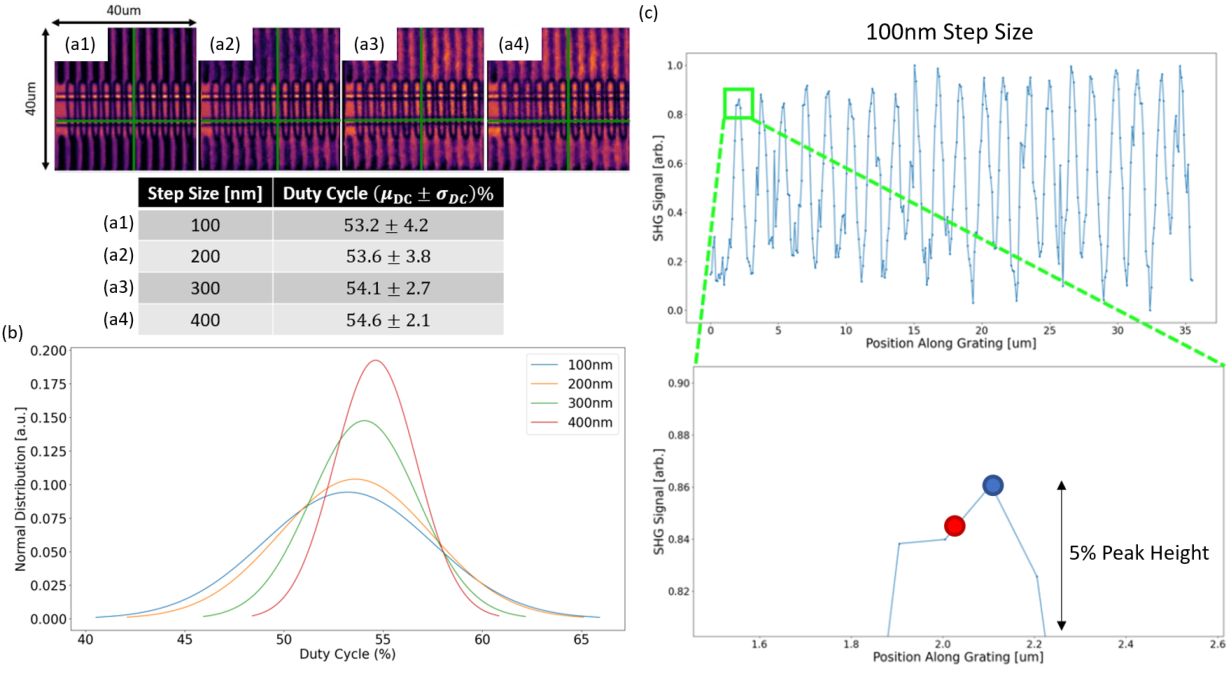}
   \end{center}
   \caption[example] 
    { \label{fig:step_size} 
    (a1-a4) illustrate a sequence of SHM images where the only change in acquisition parameters is the step size, which is swept from \qtyrange[]{100}{400}{\nm}. The overlaid, horizontal green lines display the location of SHG signal analysis. The laser power was set slightly too high, so the electrodes were damaged with repeated imaging (a2-a4). The table below summarizes the extracted mean duty cycle and duty cycle standard deviation. Normal distributions corresponding to these extracted paramters are shown in (b). Finally, (c) plots the inverted and normalized SHM signal for the  \qty[]{100}{\nm} step size data. The below zoom-in highlights the proximity of domain wall positions found with our peak-finding algorithm for the \qty[]{400}{\nm} (red) and \qty[]{100}{\nm} (blue) step size data.
    }
\end{figure}

To test our numerical models of device performance, we needed to acquire full-device-length SHM data. Our confocal beam-scanning SHM had a field of view of $\sim$\qtyproduct{100 x 100}{\um}, and the poled region of our device was \qty[]{5.6}{\mm}-long by \qty[]{15}{\um}-wide. To include the electrode fingers in the SHM image as a reference structure but avoid imaging irrelevant chip areas, we acquired a 57-image set\footnote{The first and last SHM images included the unpoled TFLN regions at the beginning and end edges of the device. So, 57 images were acquired to cover the full device area instead \qty[]{5.6}{\mm}/\qty[]{100}{\um} = 56 images.} of  \qtyproduct[]{100 x 40}{\um} images. The maximum scan speed of our galvos was \qty[]{1}{\kHz} and our image discretization ($\sim$\qtyproduct[]{400 x 400}{\nm} pixels) was set to roughly half the FWHM of our SHM point spread function (PSF). So, the whole device took $\sim$30 minutes to image, as each field took $\sim$\qty[]{30}{\second}. We note that the fastest imaging speed reported in the literature used a galvo scan speed of \qty[]{12}{\kHz} and step size of \qty[]{200}{\nm}\cite{Reitzig2021}. To validate that the increased step size of the 2D raster scan did not hamper evaluation of QPM grating quality, we took a sequential set of SHM images. The acquisition parameters and positioning of these four images were identical, except the step size was swept from \qtyrange[]{100}{400}{\nm}. The results of this systematic error characterization for a device with poling period $\Lambda = $ \qty[]{3.240}{\um} are shown in Fig. \ref{fig:step_size}. We note that only a 1\%DC difference in $\mu_{DC}$ is observed between the \qty[]{400}{\nm} and \qty[]{200}{\nm} step size measurements. Likewise, only a 2\%DC difference in $\sigma_{DC}$ is observed. The change in device performance corresponding to these perturbations in QPM grating summary metrics is negligible, so the larger step size SHM images can be used to accurately measure $\mu_{DC}$ and $\sigma_{DC}$. Because SHM images are generated using a 2D raster scan, a 2x increase in transverse step size corresponds to an overall imaging speedup of 4x.

Interestingly, we observe a decrease in $\sigma_{DC}$ as the step size increases from \qtyrange[]{100}{400}{\nm}. We can understand this by focusing on Fig. \ref{fig:step_size}\textcolor{blue}{(c)}. Rather than a perfect Gaussian curve, we see a signal plateau in the zoom-in of the \qty[]{100}{\nm} data. SHM signal fluctuation in this plateau represents a combination of noise and actual signal variation, as domain inversion depth and the presence of filamented domains can change below the diffraction-limited PSF of the SHM\cite{Rsing2019}. Using 5\% of the peak height as a threshold, the 20-point average peak width in Fig. \ref{fig:step_size}\textcolor{blue}{(a1)} is 160nm. Put another way, each peak position calculated for the \qty[]{100}{\nm} step size data has an upper bound uncertainty of $\pm$\qty[]{80}{\nm}. Propagating this error through the duty cycle calculation, which computes the difference between consecutive peak positions, the maximum systematic measurement uncertainty of our SHM and peak finding algorithm is $\Delta \text{DC} \approx4.9\%$. For QPM gratings with $\sigma_{DC} < 5\%$, this means that our measurement error is on the order of the domain width fluctuations we want to measure. Consequently, our SHM data slightly overestimates $\sigma_{DC}$ for near-ideal QPM gratings. For poor-quality QPM gratings with $\sigma_{DC}>10\%\text{DC}$, the actual variation in inverted domain shape dominates. Returning to Fig. \ref{fig:step_size}\textcolor{blue}{(c)}, the \qty[]{400}{\nm} step size SHM data is effectively a blurred copy of the  \qty[]{100}{\nm} curve shown. Consequently, the placement of peaks for the \qty[]{400}{\nm} step size trends towards the centroid of SHM signal peaks, and this placement lies within the systematic uncertainty of the \qty[]{100}{\nm} step size measurement. Thus, we can increase our SHM step size from \qtyrange[]{200}{400}{\nm} without adding significant systematic error to our measurement of $\sigma_{DC}$. 

With the full-device-length set of SHM images, we can now compute the QPM grating summary metrics $\left( \mu_{DC}, \sigma_{DC}, P_{mr} \right)$. We found that stitching SHM images together, using the dark electrode fingers as a reference geometry between images, was prone to artificial stitching errors which mar later analysis. Specifically, the coupled rate equations are highly sensitive to the phase relationship between propagating waves. Near-pixel-perfect image stitching, which is very challenging to robustly implement, is necessary to accurately model device performance. To avoid this issue, we aggregated poling pattern data from the individual SHM images to measure $\mu_{DC}, \sigma_{DC}$, and $P_{mr}$. To calculate quasi-analytic correction factors, these parameters were directly substituted into Equations \ref{eq:correction_factors}\textcolor{blue}{(a)-(c)}. 

\begin{figure} [t]
   \begin{center}
   \includegraphics[height=6cm]{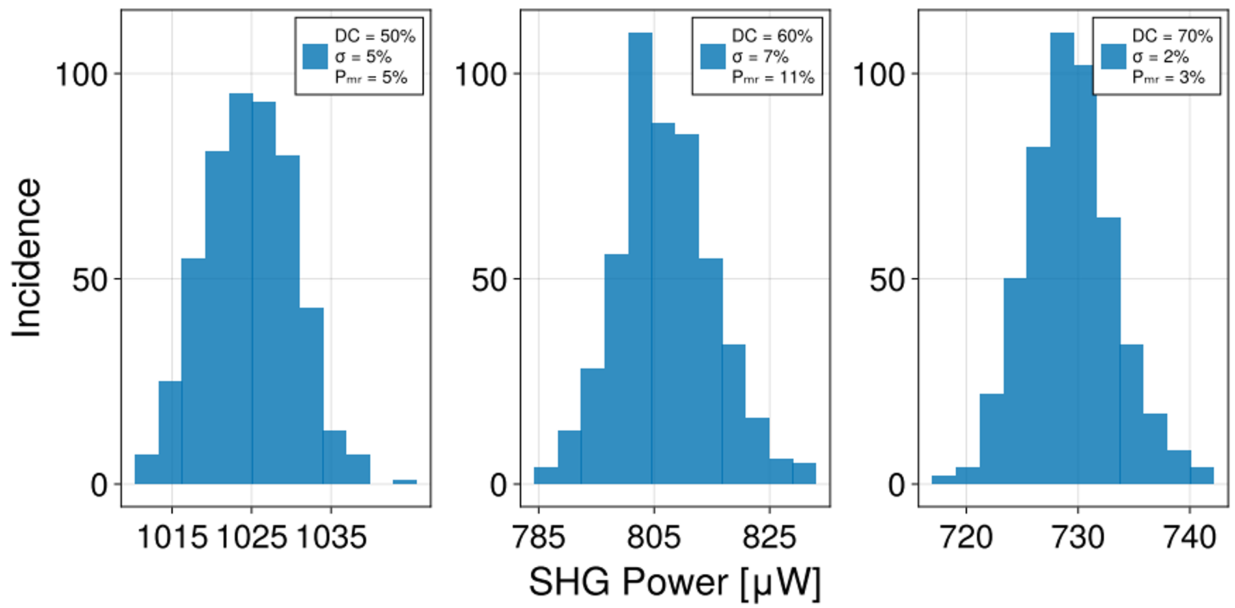}
   \end{center}
   \caption[example] 
    { \label{fig:histograms} 
    Example distributions of predicted power from N = 500 Monte Carlo simulations of the coupled rate equations. Unique QPM gratings are sampled from distributions described by fixed values of duty cycle, its variance, and the probability of missing reversals.
    }
\end{figure}

To use our coupled rate equations model without a measured and perfectly-stitched QPM grating, we instead performed Monte Carlo simulation to predict $P_{SH}$. Specifically, $N=500$ random, \qty[]{5.6}{\mm}-long $\chi^{(2)}(y)$ gratings were generated with their duty cycle distributed $\sim\!\mathcal{N}(\mu_{DC},\,\sigma_{DC}^{2})$ and missing reversals distributed $\sim\!\text{Bernoulli}(P_{mr})$. Example distributions of the predicted $P_{SH}$ for various QPM grating conditions are shown in Fig. \ref{fig:histograms}. A key observation is that the min/max spread in predicted $P_{SH}$ is only about 1-3\% of the mean. With the knowledge that our predictions of device performance will have a reasonable uncertainty bound, we can compare our numerical models with experimental results.

Using the full set of 57 SHM images of the device shown in Fig. \ref{fig:DeviceStructure}\textcolor{blue}{(a)}, $\mu_{DC}, \sigma_{DC},$ and $P_{mr}$ were extracted. Using these parameter values, we calculated $\eta_{DC}, \eta_{\sigma_{DC}},$ and $\eta_{mr}$. Using the propagation losses at the peak second harmonic wavelength (\qty[]{789}{\nm}) and corresponding fundamental, c-band wavelength (\qty[]{1578}{\nm}), we calculated $\eta_{prop}$. The conversion efficiency for an ideal device (lossless and perfect 50\%DC square wave QPM grating) was simulated using Equations \ref{eq:coupled}\textcolor{blue}{(a)-(c)}. Finally, the empirical device conversion efficiency was computed by measuring the coupling losses, pump power, and second harmonic output power. Coupling losses were used to compute the in-waveguide pump power and $P_{SH}$. The end result of these computations is the link budget shown in Tab. \ref{tab:link_budget}. Within experimental uncertainty, the predicted and measured values of the in-waveguide second harmonic power $(P_{SH})$ and length-normalized conversion efficiencies $(\eta)$ are in agreement.

\begin{table}[ht]
    \caption{Link budget calculation to check agreement between predictions and measurements of device performance. The uncertainty in the corrected conversion efficiency is dominated by the propagation loss and duty cycle measurements.} 
    \label{tab:link_budget}
    \begin{center}       
    \begin{tabular}{l | c}
    \textbf{Link Factor} & \textbf{Value} \\
    \hline
    Ideal conversion efficiency $\eta_{0}$ & \qty[]{1610}{\percent \per \watt \per \cm^2} \\
    Duty cycle mean correction $\eta_{DC}$ & 0.88(2) \\
    Duty cycle variation correction $\eta_{\sigma_{DC}}$ & 0.98 \\
    Missing reversal correction $\eta_{mr}$ & 0.94 \\
    Propagation loss correction $\eta_{prop}$ & 0.32(4) \\ 
    \hline
    Corrected conversion efficiency $\eta$ & \qty[]{420(60)}{\percent \per \watt \per \cm^2} \\
    $\rightarrow$ Predicted SH power & \qtyrange[]{16.8}{22.6}{\uW} \\
    \hline
    Device length & \qty[]{0.56}{\cm} \\
    In-waveguide pump power & \qty[]{3.875}{\mW} \\
    \hline
    In-waveguide SH power & \qty[]{22.1}{\uW} \\
    Measured conversion efficiency $\eta$ & \qty[]{470}{\percent \per \watt \per \cm^2}  \\
    \hline 
    \end{tabular}
    \end{center}
\end{table}

\begin{figure} [b]
   \begin{center}
   \includegraphics[height=6cm]{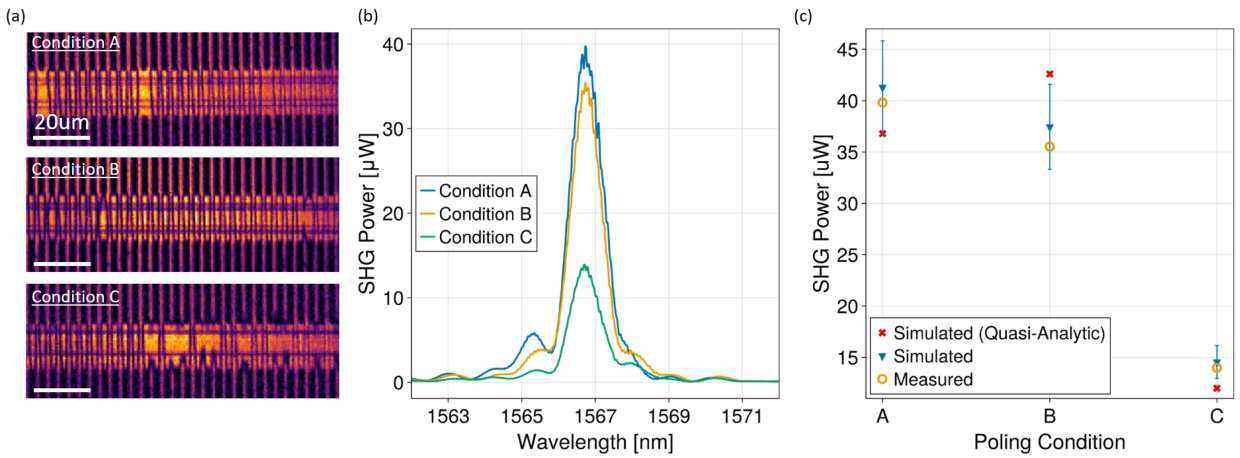}
   \end{center}
   \caption[example] 
    { \label{fig:repoling} 
    (a) Shows representative SHM images of the different experimental poling conditions. For each condition, a total of 57 SHM images were taken to cover the full length of the device. (b) Plots the QPM spectra measured for each poling condition. (c) Compares the measured SH power (orange) to predictions based on quasi-analytic correction factors (red) and solving the coupled rate equations (blue). The error bars for the rate equation approach correspond to the $1\sigma$ spread from 500 Monte Carlo simulations of the SH power. 
    }
\end{figure}

The link budget shown in Tab. \ref{tab:link_budget} is for a device with a high-quality QPM grating. Often times, it can be challenging to achieve near-ideal poling. So, we wanted to test our hypothesis---device performance can be accurately predicted using quasi-analytic correction factors and our coupled rate equation model---for poorer-quality QPM gratings. The experimental setup and results are shown in Fig. \ref{fig:repoling}. A duplicate of the device shown in Fig. \ref{fig:DeviceStructure}\textcolor{blue}{(c)}, which was on the same chip and poled with the same parameters, was used for this experiment. The QPM grating quality after the initial poling pulse corresponds to Condition A, and a representative image is shown in Fig. \ref{fig:repoling}\textcolor{blue}{(a)}. The peak SH output power of the device and QPM spectrum were measured. Next, the device was repoled at \qty[]{200}{\degreeCelsius} with the same electrical pulse parameters: peak voltage $V_{pk} = $ \qty{475}{\volt}, hold time at peak voltage $t_{hold} = $ \qty{80}{\ms}, and ramp/fall rates of $500/100$ \unit[per-mode = symbol]{\volt\per\ms}. At Condition B, the QPM spectrum and peak SH output power of the device were again measured. We found that applying consecutive pulses has a diminishing impact on domain inversion. This is likely due to the presence of conductive domain walls which promote the injection of a leakage current whose charge carriers screen and locally reduce the applied poling E-field\cite{Nagy2019, Doshi2025}. Therefore, we heated the device to \qty[]{200}{\degreeCelsius} and applied an additional two high voltage pulses with $V_{pk}$ increased to \qty[]{600}{\volt}; this promoted further degradation of the QPM grating through exaggerated domain broadening and merging. At Condition C (and for the last time), the peak second harmonic output power and QPM spectrum of the device were measured.

The QPM spectra for the different conditions are shown in Fig. \ref{fig:repoling}(b). As expected, we see the peak of the sinc\textsuperscript{2} shape decrease as the QPM grating quality decreases. The small shoulders on each side of the main lobe are characteristic of 1/f-like (spatial frequencies) phase mismatch variations along the length of the waveguide\cite{Helmfrid91}. Such inhomogeneous broadening reduces the peak efficiency\textemdash here, the effect is on the order of a few percent and we neglect it. 
Fig. \ref{fig:repoling}(c) compares the measured value of $P_{SH}$ to the predictions of $P_{SH}$ from both the quasi-analytic correction factor and rate equation models. Though non-extreme metrics parameterize the QPM grating in conditions B and C, the quasi-analytical expressions struggle to accurately predict $P_{SH}$. In contrast, we find excellent agreement between measurements of $P_{SH}$ and the rate equation predictions for all conditions.

\begin{figure} [t]
   \begin{center}
   \includegraphics[height=8.5cm]{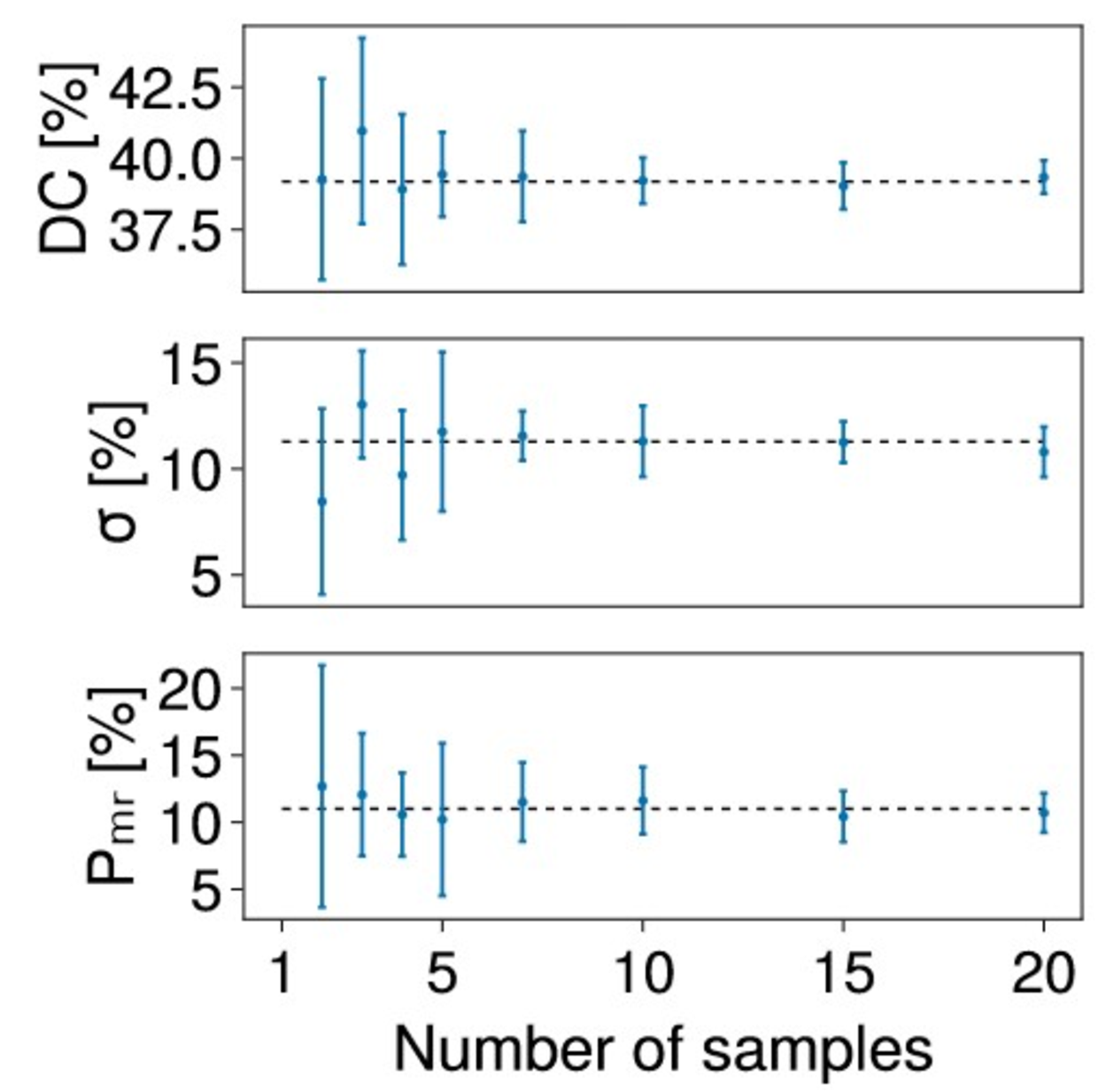}
   \end{center}
   \caption[example] 
    { \label{fig:subsampling} 
    Illustrates the effect of subsampling on estimating QPM grating quality metrics. Each point and its error bars (95\% confidence interval) are generated by repeatedly sampling with replacement the number of SHM images (from the total pool of 57 SHM images) indicated on the x-axis. For each point, sampling with replacement is repeated 20 times. 
    }
\end{figure}

In the process of comparing our models with experimental second harmonic generation data, we realized that aggregating individual SHM images to estimate $\mu_{DC}, \sigma_{DC},$ and $P_{mr}$ is analogous to sampling from a statistical population. Barring any mm-scale poling quality correlations, which we never observed in our experiments, we hypothesized that only a small sample of the QPM grating periods is needed to get an accurate picture of the poling pattern distribution.  Fig. \ref{fig:subsampling} shows the progression of aggregating data from an increasing number of SHM images, sampled with replacement at random from the complete set of 57, to estimate QPM grating summary metrics. Focusing on the error bars, which indicate a 95\% confidence interval, we find that a sample of just 10 SHM images is sufficient to compute quality metrics to within 1\% of their asymptotic (full-device-length) value. 10x\qty[]{100}{\um}-long SHM images corresponds to $\sim$300 periods, which for this device is roughly 20\% of the QPM grating. Thus, lengthwise subsampling can provide a 5x SHM imaging speedup. We also note that this method provides another reason to avoid image stitching---the optimal way to ensure detection of any potential mm-scale poling pattern correlations would be to uniformly space the SHM images across the length of the waveguide. By combing both QPM grating length and SHM scan discretization subsampling, we can achieve an order of magnitude speedup. Moreover, these SHM images can be used to accurately predict device performance for a wide variety of QPM grating quality conditions using Monte Carlo simulation of the coupled rate equations.

\section{Discussion}
\label{sec:disc}  
In Fig. \ref{fig:repoling}, we found excellent agreement between rate equation simulations and measured SH power. We note that highly-optimized stitching algorithms could enable the direct input of a measured QPM grating into the nonlinear coupling coefficients $\kappa_i$. In lieu of this approach, Monte Carlo simulations using synthetic gratings were able to predict output SH power with a reasonably narrow distribution ($\sim$1-3\% of the mean value). We note that our choice to neglect the correction factor for inhomogeneous broadening\cite{Helmfrid91} and the $\sim$1-2\% systematic error in our measurement of duty cycle variation $(\sigma_{DC})$ may account for the slight overestimation of measured powers in Fig. \ref{fig:repoling}\textcolor{blue}{(c)}. However, this correction does not account for the disagreement between the predicted value of $P_{SH}$ using quasi-analytic correction factors and the measured value of $P_{SH}$ for conditions B and C. Intuitively, the failure of the correction factors makes sense as the QPM approximation\cite{Boyd} (perfect 50\% DC square wave modulation of $\chi^{(2)}$) breaks down. Higher order terms in the Fourier series decomposition become significant, and the expression for $\eta_{DC}$ in particular becomes inaccurate. While these correction factors are still useful for building intuition and validating near-optimal device performance, the rate equation approach is necessary for degraded QPM grating quality. Overall, both methods are valuable for process optimization, as they can reveal sources of non-ideal device performance.

The rapid convergence of $\mu_{DC}$, $\sigma_{DC}$, and $P_{mr}$ shown in Fig. \ref{fig:subsampling} suggests that a subsample of only $N=300$ QPM grating periods is sufficient to measure summary statistics to within 1\% of their full-device values. In our experiments, domain inversion quality did not appear to be broadly spatially correlated, suggesting that the process is insensitive to parameters such as TFLN thickness, which varies a few nm over mm-scale lengths\cite{Chen2023, Zhao2023}. For the device characterized, a sample of 300 periods corresponds to $\sim$20\% of total grating length. However, we expect that increasing the length of the QPM grating would not require a larger sampling set to estimate the grating parameters. We note that the number of samples required should, in extreme cases, increase as domain inversion quality degrades significantly. For fabrication processes or devices which exhibit large non-uniformity along the waveguide length, additional characterization may be necessary. For average to good-quality QPM gratings, though, in-line process metrology for determining device quality can extract representative QPM grating summary metrics with a 5x imaging speedup. 

We note that the device imaged in this work had a poling period of $\Lambda = $ \qty[]{3.240}{\um}. Thus, the 4x imaging speedup from increasing the step size from \qtyrange[]{200}{400}{\nm} can be readily generalized to devices with longer poling periods. However, QPM gratings with shorter poling periods may necessitate smaller step sizes to accurately capture grating quality. Moreover, narrower SHM PSFs (or in extereme cases, PFM) may be needed for QPM gratings with submicron poling periods\cite{Zhao2020SubMicron}. In contrast, subsampling of the poling pattern by imaging only segments of the total device length can be applied to any arbitrary QPM grating. Increased SHM resolution can be achieved through the use of higher numerical aperture (NA) objectives, but there may be optical system tradeoffs. Because high-NA objectives typically have large magnifications, their field of view is reduced. This can cause significant imaging slowdown, as acquiring additional fields requires sample repositioning and refocusing. Moreover, NA$>$1.0 objectives require the use of an immersion medium, which may not be appropriate for sensitive samples. On the flip side, lower magnification objectives could be used for larger field of view imaging. Unfortunately, the associated reduction in NA reduces the measurement signal-to-noise ratio (SNR) \cite{Rsing2019}. Compensating for reduced SNR via increasing the average power on sample (for a fixed pulse width) is not always possible, as this risks damaging the sample. Alternatively, the integration time of the SHM detection scheme can be increased, but this incurs an imaging speed penalty. These tradeoffs can be used to optimize a SHM instrument for a particular task, or they can enable a setup to characterize a wide variety of material samples. The flexibility to tackle multi-scale characterization makes SHM a robust imaging modality.

A number of groups are working on scaling TFLN fabrication from chip-scale to wafer-scale\cite{Luke20,Chen2024}. To that end, we anticipate that combining subsampling with the enhanced imaging speed of an industrial SHM\cite{Reitzig2021} will be critical for future, wafer-scale characterization of TFLN devices. With the fastest SHM imaging speed of TFLN reported in the literature (\qty[per-mode=symbol]{1}{\cm^2\per} hour), only $\sim$0.5 wafers (100mm)/day could be imaged. With step size and poling pattern subsampling, up to 10 wafers a day could be characterized! Along the theme of scalability, we envision an idealized wafer probing setup for TFLN and more generally, scalable materials characterization. The setup would have electrical probing ports for DC, high voltage, and RF experiments. Optical probing ports could be used to analyze grating-coupled devices. A joint, motorized piezo and large-travel stage would be used in combination with a brightfield/darkfield microscope for fine and coarse navigation. Finally, a confocal beam-scanning port would multiplex SHM with other modalities such as broadband reflectometry\cite{Chen2023} or Raman microscopy\cite{Reitzig2021}. Such a robust, multimodal setup would be invaluable for investigating a wide variety of materials and realizing the full potential of the TFLN platform.

\section{Conclusion}
\label{sec:conc}  
We used SHM to iteratively optimize periodic poling of a hybrid SiN-on-TFLN device. We used this device to show that SHM images of the QPM grating, in conjunction with numerical models, can predict device performance. Specifically, quasi-analytic correction factors that account for deviation of the mean duty cycle from 50\%, variation in the duty cycle, missing reversals, and propagation losses are only effective when the overall quality of the QPM grating is high. In contrast, Monte Carlo simulations of the coupled rate equations accurately predict device performance for a wide range of QPM grating qualities. By increasing the step size of our SHM raster scan by 2x, we were able to achieve a 4x imaging speedup without compromising the accuracy of our QPM grating analysis. Moreover, a subsample of only 300 periods was sufficient to measure QPM grating summary metrics to within 1\% of their full-device value. 300 periods corresponded to $\sim$20\% of the total QPM grating length, so SHM imaging can proceed an additional 5x faster. By combing discretization and poling pattern length subsampling, we can acquire SHM images an order of magnitude faster. These results underscore that SHM is a powerful tool for process and poling optimization. The presented framework for QPM grating analysis using SHM can be generalized to other scalable materials characterization challenges, and it is an important step towards realizing the full potential of the TFLN platform.
\acknowledgments 
 
We would like to thank collaborators at MIT Lincoln labs for their input. Moreover, we would like to thank our sponsors, as this work was supported by the DARPA DSO Robust Optical Clock Network (ROCkN) Program (FA8702-18-F-0004) and in part by an appointment to the Intelligence Community Postdoctoral Research Fellowship Program at the Massachusetts Institute of Technology (ICPD-2019-01). This fellowship is administered by the Oak Ridge Institute for Science and Education (ORISE) through an interagency agreement between the U.S. Department of Energy and the Office of the Director of National Intelligence (ODNI). This work was carried out in part through the use of MIT.nano's facilities.

\bibliography{report} 
\bibliographystyle{spiebib} 

\end{document}